\documentclass[aps,prd,twocolumn,superscriptaddress,amssymb,eqsecnum,showpacs,showkeyes,secnumarabic,graphics,floatfix,nofootinbib,tightenlines,longbibliography]{revtex4-1}
\usepackage{relsize}
\usepackage{blindtext}
\usepackage{enumitem}
\usepackage{xcolor}
\usepackage{graphicx} 
\usepackage{epsfig}
\usepackage{bm}
\usepackage{cancel}
\usepackage{dcolumn}
\usepackage{amsmath}
\usepackage{hhline} 
\usepackage[utf8]{inputenc}
\usepackage[breaklinks=true,colorlinks=true,
linkcolor=blue,urlcolor=blue,citecolor=cyan,% PDF VIEW
bookmarks=true,bookmarksopenlevel=2]{hyperref}

\def\apj{{Astroph.\@ J.\ }}

\def \beq  {\begin{equation}}
\def \eeq  {\end{equation}}
\def \ber  {\begin{eqnarray}}
\def \eer  {\end{eqnarray}}

\begin{document}
\newcommand{\newc}{\newcommand}

\newc{\be}{\begin{equation}}
\newc{\ee}{\end{equation}}
\newc{\ba}{\begin{eqnarray}}
\newc{\ea}{\end{eqnarray}}
\newc{\bea}{\begin{eqnarray*}}
\newc{\eea}{\end{eqnarray*}}
\newc{\D}{\partial}
\newc{\ie}{{\it i.e.} }
\newc{\eg}{{\it e.g.} }
\newc{\etc}{{\it etc.} } 
\newc{\etal}{{\it et al.}}
\newc{\lcdm}{$\Lambda$CDM }
\newcommand{\nn}{\nonumber}
\newc{\ra}{\Rightarrow}

\title{Reconstructing a Model for Gravity at Large Distances from Dark Matter Density Profiles}
\author{L. Perivolaropoulos}\email{leandros@uoi.gr} 
\author{F. Skara}\email{fskara@cc.uoi.gr}
\affiliation{Department of Physics, University of Ioannina, 45110 Ioannina, Greece}

\date {\today}  

\begin{abstract}
Using the Navarro-Frenk-White (NFW) dark matter density profile we reconstruct an effective field theory model for gravity at large distances from a central object by demanding that the vacuum solution has the same gravitational properties as the NFW density profile has in the context of General Relativity (GR). The dimensionally reduced reconstructed action for gravity leads to a vacuum metric that includes a modified Rindler acceleration term in addition to the Schwarzschild and cosmological constant terms. The new term is free from infrared curvature singularities and leads to a much better fit of observed galaxy velocity rotation curves than the corresponding simple Rindler term of the Grumiller metric \cite{Grumiller:2010bz,Grumiller:2011gg}, at the expense of one additional parameter. When the new parameter is set to zero the new metric term reduces to a Rindler constant acceleration term. We use galactic velocity rotation data to find the best fit values of the parameters of the reconstructed geometric potential and discuss possible cosmological implications.
\end{abstract}
\maketitle  
  
\section{Introduction}   
\label{sec:Introduction}    

General Relativity (GR) is the simplest successful theory for gravity. It is consistent with the vast majority of experiments and observations from sub-mm scales up to cosmological horizon scales \cite{Fischbach:1999bc,Will:2005va}. Alternative theories of gravity include more degrees of freedom and parameters which are strongly constrained by a wide range of experiments and astrophysical/cosmological observations to be very close to the values predicted by GR (see e.g. \cite{Kapner:2006si,DiPorto:2007ovd,Nesseris:2007pa,Daniel:2010ky}).   
  
Despite of its successes and simplicity, GR requires additional undetected matter/energy components to explain observations on galactic scales or larger. In particular, the existence of dark matter \cite{1933AcHPh...6..110Z,1937ApJ....86..217Z,1970ApJ...160..811F,Rubin:1970zza,Rubin:1980zd,Bosma:1981zz,Bertone:2004pz} is required for the description of observed dynamics and structure formation on galactic scales or larger while dark energy with negative pressure or a fine tuned cosmological constant (see Ref. \cite{Peebles:2002gy} for a review) is required for the consistency of GR with the observed accelerating cosmic expansion \cite{Riess:1998cb,Riess:1998dv,Perlmutter:1998np}.  Even on solar system scales or sub-mm scales there have been hints of possible inconsistency of the theory with particular observations (e.g. Pioneer anomaly \cite{Anderson:1998jd,Anderson:2001sg,Dittus:2005re,Lammerzahl:2006ex,Turyshev:2012mc}) or short range gravity experiments (peculiar oscillating signals in some datasets \cite{Perivolaropoulos:2016ucs,Antoniou:2017mhs}). In addition, the theory predicts the existence of unphysical singularities in a wide range of its solutions which should describe physical phenomena.   
       
Any observed inconsistency between the geometric left hand side (LHS) of the Einstein equation and the matter-energy right hand side (RHS) is thus usually addressed by modifying the RHS through the conservative assumption of some yet undetected form of matter-energy chosen in such a way as to restore the equality of the geometric and matter parts of the Einstein equation. A more fundamental approach is to modify the geometric LHS of the Einstein which is equivalent to modifying the fundamental action of the gravitational theory. There is a wide range of modified gravity models aiming at the explanation of the accelerating expansion of the universe \cite{Copeland:2006wr,Frieman:2008sn,Caldwell:2009ix,Clifton:2011jh}. Such theories include scalar tensor theories \cite{Brans:1961sx,Uzan:1999ch,Boisseau:2000pr,Schimd:2004nq} including the most general class of Horndeski models \cite{Horndeski:1974wa,Charmousis:2011bf}, $f(R)$ theories \cite{Starobinsky:1980te,Hu:2007nk,Fay:2007uy,DeFelice:2010aj,Nojiri:2010wj,Basilakos:2013nfa} which generalize the Ricci scalar $R$ of the action to a general function $f(R)$, generalized teleparallel gravity $f(T)$ theories \cite{Ferraro:2006jd,Ferraro:2008ey,Linder:2010py,Nesseris:2013jea} which generalize the torsion scalar $T$ of the action to a general function of it, non-local gravity theories \cite{Mashhoon:1993zz,Mashhoon:2008vr,Deser:2007jk} which introduce nonlocal operators in the gravitational action which involve effectively an infinite sum of derivatives etc. On the other hand, modified gravity models aiming at the explanation of the dynamics of matter at galactic and cluster scales without dark matter are much more limited \cite{Capozziello:2004us,Martins:2007uf}. This is due to the very diverse nature of matter dynamical behaviors that need to be explained which appears to require a large number of parameters for the fundamental theory that would attempt to explain it without dark matter. The main representative of this class of theories is the modified Newtonian Dynamics (MOND) theory \cite{Milgrom:1983ca,Sanders:2002pf,Famaey:2011kh} based on the existence of a fundamental acceleration scale which has been recently shown however to be highly unlikely to exist \cite{Rodrigues:2018duc}.      
   
An alternative approach towards a geometric fundamental description of the dynamics of matter on galactic and cluster scales without dark matter has been proposed by Grumiller \cite{Grumiller:2010bz,Grumiller:2011gg}. Assuming spherical symmetry of the metric and implementing dimensional reduction of the Einstein Hilbert action to two space-time dimensions ($t-r$) it was shown that the emerging 2-dimensional scalar tensor effective field theory action with a constant potential can be generalized to include a non-trivial potential. The simplest form of this potential with no infrared curvature singularities, leads to a generic Rindler constant acceleration term in the vacuum spherically symmetric metric of the new theory \cite{Grumiller:2010bz}. It has been shown recently \cite{Alestas:2019wtw} that such a term in the background metric can give rise to a new type of metastable topological defects (spherical domain walls).  It was also argued that such a term can give rise to the observed velocity rotation curves of galaxies without incorporating dark matter \cite{Lin:2012zh}. It was later shown \cite{Mastache:2012ep,Cervantes-Cota:2013hza} however that the Rindler term is only able to provide acceptable fits to a relatively small number of observed velocity rotation curves which is limited to those rotation curves where the velocity continues to increase with distance through the halo. Such a behavior is not typical for most rotation curves which are either flat \cite{Rubin:1970zza,Rubin:1980zd,Bosma:1981zz} or in fact tend to decrease with distance at large distances from the galactic core \cite{Salucci:2007tm}. Thus, the Rindler acceleration even though it is appealing due to its possible fundamental geometric origin, does not provide enough degrees of freedom to describe the data in contrast to the commonly used dark matter density profiles (Navarro-Frenk-White \cite{Navarro:1995iw,Navarro:1996gj} and Burkert \cite{Burkert:1995yz}) which provide excellent fits to the rotation curve data. Thus, the following questions arise:    
\begin{itemize}    
\item
{\it Is it possible to generalize the fundamental 2-dimensional geometric effective action (and its scalar field potential emerging from dimensional reduction) such that the corresponding vacuum spherically symmetric metric reproduces the observed velocity rotation curves equally well as the standard dark matter density profiles?}  
\item
{\it If yes, what is the form of the required geometric scalar field potential and how is it related to the simple Rindler potential of Refs. \cite{Grumiller:2010bz,Grumiller:2011gg}}  
\item
{\it Can an arbitrary vacuum spherically symmetric metric be reproduced by a properly selected geometric scalar field potential?}
\end{itemize}
The goal of the present analysis is to address these questions using both theoretical reconstruction of the fundamental action and direct comparison with specific velocity rotation data. 

The structure of this paper is the following: In the next section we consider a class of simple spherically symmetric metrics in $3+1$ dimensions and identify the profiles and properties of the perfect fluids that can give rise to such metrics. In section III we assume spherical symmetry and use it to dimensionally reduce the $3+1$ dimensional Einstein-Hilbert action to an effective two dimensional scalar-tensor action with a constant potential. We generalize this geometric potential thus modifying the gravitational action to an arbitrary form and derive the corresponding generalized vacuum spherically symmetric metric in terms of the geometric potential.  In section IV we consider special forms of the geometric potential and of the background fluid and derive the corresponding metric. Thus the case of a constant potential (GR) we derive the Schwarzschild vacuum metric while for a simple quadratic potential we obtain the Rindler acceleration and cosmological constant terms in agreement with Ref.  \cite{Grumiller:2010bz}. We also reconstruct the geometric potential that leads to a vacuum metric that is identical with the metric derived assuming a given dark matter fluid density profile in the context of GR. In the context of a particular example we assume a Navarro-Frenk-White (NFW) \cite{Navarro:1995iw,Navarro:1996gj} density profile and derive the corresponding geometric potential and vacuum metric. We show that this metric generalizes the Rindler term of the Grumiller metric and show fits of the velocity profiles it generates on typical galactic velocity rotation data. In what follows we assume a metric signature $+---$.

\section{Spherically Symmetric Metrics in GR and Perfect Fluids}
\label{sec:Spherically Symmetric Metrics}

Consider the spherically symmetric metric in 4-dimensions of the form
\be
ds^2=f(r)dt^2-f(r)^{-1}dr^2-r^2(d \theta^2 +sin^2 \theta d \phi^2)
\label{eq:metric4d} 
\ee
{\it What is the most general form of perfect fluid energy momentum tensor that is consistent with this metric in the context of GR?}

In order to address this question we set
\be
f(r)=1-g(r)
\label{eq:fg}
\ee
and obtain the Einstein tensor corresponding to this metric as  
\be
G_{\mu} ^{\nu}= \begin{bmatrix}
e_1(r)&0&0&0\\
0&e_1(r)&0&0\\
0&0&e_2(r)&0\\
0&0&0&e_2(r)
\end{bmatrix}  
\label{eq:gmn}  
\ee 
with
\ba
e_1(r)&=&\frac{g(r)}{r^2} + \frac{g'(r)}{r}
\label{eq:ee1}\\
e_2(r)&=&\frac{g'(r)}{r} +\frac{g''(r)}{2}
\label{eq:ee2}
\ea
Using eqs.(\ref{eq:ee1}), (\ref{eq:ee2})  and the Einstein equations $G^{\mu}_{\nu} = \kappa T^{\mu}_{\nu}$ we find
\ba  
\rho(r)&=&-p_r(r)=\frac{1}{\kappa r}[\frac{g(r)}{r}+g'(r)]\\
p_\theta(r)&=&p_\phi(r)=-\frac{1}{2\kappa r}[2g'(r)+rg''(r)]
\ea
where $\kappa=8\pi G$ and the energy momentum tensor of the perfect fluid is
\be 
 T_\mu^\nu=diag\left[\rho(r),-p_r(r),-p_\theta(r),-p_\phi(r)\right] 
\ee\\

Expanding $g(r)$ as a power series\\
\be 
f(r)=1-\sum^{N}_{n=-N}a_{n}r^{n}
\label{eq:fpoly}
\ee

the Einstein tensor may be expressed as \cite{Alestas:2019wtw}
 
\begin{widetext}
\be
G^{\mu}_{\nu}=\sum^{N}_{n=-N}
  \begin{bmatrix}
    a_{n}(n+1)r^{n-2} & 0 & 0 & 0 \\
    0 & a_{n}(n+1)r^{n-2} & 0 & 0 \\
    0 & 0 & \frac{1}{2}a_{n}n(n+1)r^{n-2} & 0 \\
    0 & 0 & 0 & \frac{1}{2}a_{n}n(n+1)r^{n-2}
  \end{bmatrix}
\ee
\end{widetext}
Therefore, the energy - momentum tensor supporting the metric function (\ref{eq:fpoly}) is  
\begin{align}
T^{0}_{0} &=\frac{1}{\kappa}\sum^{N}_{n=-N}a_{n}(1+n)r^{n-2}=\rho \label{rho}\\
T^{r}_{r} &=T^{0}_{0}=-p_{r} \label{p_r}\\
T^{\theta}_{\theta} &=\frac{1}{2\kappa}\sum^{N}_{n=-N}a_{n}n(1+n)r^{n-2}=-p_{\theta} \label{ptheta}\\
T^{\phi}_{\phi} &=T^{\theta}_{\theta}=-p_{\phi} \label{pPhi} 
\end{align} 
 
 As expected the term $n=-1$ (Schwarzschild metric) corresponds to the vacuum solution ($\rho=p=0$) while for $n=2$ we have the cosmological constant term (constant energy density-pressure). The Rindler constant acceleration term $n=1$ is generated by a perfect fluid with 
\be  
 \rho=\frac{2 a_1}{\kappa r}=-p_r=-2 p_\theta=-2 p_\phi
\label{rindfluid}
\ee 
For $n=0$ (constant term in the metric function) we have the case of a global monopole (zero angular pressure components and energy density $\sim r^{-2}$ \cite{Barriola:1989hx,Shi:2009nz,Bennett:1990xy,Harari:1990cz,Dadhich:1997mh}). Thus any power law term of the spherically symmetric metric function $g(r)$ can be generated by a corresponding power law term of the energy momentum tensor of a perfect fluid provided that its radial pressure equation of state parameter $w_r$ is $-1$ and there is equality between the angular pressure components. 
 
  The question we address in the next section is the following: {\it Can the spherically symmetric metric (\ref{eq:metric4d}) also emerge as a vacuum solution in a modified gravity theory?} In other words, given a spherically symmetric fluid and its corresponding metric in the context of GR, what is the spherically symmetric modified gravity theory that leads to the same metric as its vacuum solution?
 
 \section{Modifying Spherically Symmetric GR through Dimensional Reduction }
\label{sec:Dimensional Reduction}
Consider the generalization of the  spherically symmetric metric (\ref{eq:metric4d}) to a $d$-dimensional form
\be
ds^2=f(r)dt^2-f(r)^{-1}dr^2-\Phi(r)^2 d \Omega 
\label{eq:metricdd}
\ee
where $\Phi(r)$ denotes the the surface radius and $d\Omega$ is the solid angle in $d-2$ dimensions. The Einstein-Hilbert gravitational action describing the dynamics of the metric (\ref{eq:metricdd}) in the context of GR is of the form

\be
 S= \frac{1}{2\kappa_d}\int d^dx \sqrt{-g^{(d)}} R^{(d)} + \int d^dx \sqrt{-g^{(d)}}  {\cal L}_M^{(d)}
\label{eq:actiond}  
\ee
\\
where $R^{(d)}$ is the Ricci scalar in $d$ dimensions and ${\cal L}_M^{(d)}$ is the matter Lagrangian density assumed to describe a spherically symmetric perfect fluid. It is straightforward to show using the metric (\ref{eq:metricdd}) that the $d-$dimensional Ricci scalar can be expressed in terms of the corresponding 2-dimensional ($t-r$) scalar as \cite{Grumiller:2002nm} 
\be
R^{(d)}=R^{(2)}-\frac{(d-2)(d-3)}{\Phi^2}\left[1+(\partial\Phi)^2\right]-\frac{2(d-2)}{\Phi}\nabla^b\partial_b\Phi 
\label{eq:Rd} 
\ee
while for the $d$-dimensional spherically symmetric metric determinant we have
\be
\sqrt{-g^{(d)}}=\Phi^{d-2}\sqrt{-g^{(2)}}   
\label{eq:gd} 
\ee
Using eqs (\ref{eq:Rd}) and (\ref{eq:gd}) in (\ref{eq:actiond}) we may intergate trivially over the angular coordinates and dimensionally reduce this action to a 2-dimensional ($t-r$) scalar-tensor action of the form
\begin{widetext}  
\be
S=\frac{V_{d-2}}{2\kappa_d}\int d^2x\sqrt{-g^{(2)}}\left[\Phi^{d-2}R^{(2)}+(d-2)(d-3)\Phi^{d-4}(\partial\Phi)^2-(d-2)(d-3)\Phi^{d-4}\right]+V_{d-2}\int d^2x \sqrt{-g^{(2)}} {\cal L}_M^{(2)} 
\label{eq:redactiond} 
\ee
\end{widetext}
where $V_{d-2}$ is the $d-2$ dimensional angular volume which is equal to $4\pi$ for $d=4$.  For $d=4$ the $2$-dimensional action takes the form \\
\be
S=\frac{1}{4G}\int d^2x\sqrt{-g^{(2)}}\left[\Phi^2 R^{(2)}+2(\partial\Phi)^2-2\right]+S_M^{(2)}  
\label{eq:redaction1} 
\ee\\
A modification of spherically symmetric GR can be implemented at this stage by generalizing the effective dimensionally reduced GR action (\ref{eq:redaction1}) to a general scalar tensor action \cite{Russo:1992yg,Odintsov:1991qu} of the form
\begin{widetext} 
\be
S=\frac{1}{4G}\int d^2x\sqrt{-g^{(2)}}\left[F(\Phi) R^{(2)}-Z(\Phi)(\partial\Phi)^2-2V(\Phi)\right]+S_M^{(2)}  
\label{eq:redaction2} 
\ee  
\end{widetext}
where $F(\Phi)$, $Z(\Phi)$, $V(\Phi)$ are arbitrary functions of the field $\Phi$\footnote{Note that for the dimensionally reduced metric $\Phi(r)$ can be considered as a scalar field (up to a dimensionful parameter) in correspondence with e.g. the radion field which is an effective scalar field in 4-dimensions describing the dynamics of extra dimensions in a cosmological setup \cite{Perivolaropoulos:2002pn}  in the context of an effective scalar tensor theory in 4-dimensions.}

The origin of this generalized scalar tensor action (\ref{eq:redaction2}) could either come from physics at the effective 2-dimensional $(t-r)$ level or could emerge through dimensional reduction of a spherically symmetric scalar tensor theory.

In particular consider the $d$-dimensional scalar-tensor action 
\begin{widetext}
\be
S=\frac{1}{2\kappa_d}\int d^dx \sqrt{-g^{(d)}}\left[\chi(\Phi) R^{(d)} -\zeta (\Phi) (\partial\Phi)^2-U(\Phi)\right]+S_M^{(d)} 
\label{eq:actiondd}  
\ee  
\end{widetext} 
which for $\chi(\Phi)=1$ , $\zeta(\Phi)=0$  and $U(\Phi)=0$  reduces to the Einstein - Hilbert action (\ref{eq:actiond}). It is straightforward to show that the action (\ref{eq:actiondd}) can be dimensionally reduced using spherical symmetry and the metric (\ref{eq:metricdd}) to the 2-dimensional action 
\begin{widetext} 
\be
\begin{matrix}
S=\frac{V_{d-2}}{2\kappa_d}\int d^2x\sqrt{-g^{(2)}}\{\chi(\Phi)\Phi^{d-2}R^{(2)}+[(d-2)(d-3)\chi(\Phi)\Phi^{d-4}+2(d-2)\chi'(\Phi)\Phi^{d-3}\\
\\
-\zeta(\Phi)\Phi^{d-2}](\partial\Phi)^2-(d-2)(d-3)\chi(\Phi)\Phi^{d-4}-\Phi^{d-2}U(\Phi)\}+S_M^{(2)} 
\end{matrix}  
\label{eq:redactiond} 
\ee
where the prime ($'$) denotes derivative with respect to the surface radius field $\Phi$. Clearly for $d=4$ the action (\ref{eq:redactiond}) reduces to (\ref{eq:redaction2}) by setting
\be
F(\Phi)=\chi(\Phi)\Phi^2
\ee
\be
Z(\Phi)=-2\chi(\Phi)-4\chi'(\Phi)\Phi+\zeta(\Phi)\Phi^2 
\ee
\be 
V(\Phi)=\chi(\Phi)+\frac{\Phi^2}{2}U(\Phi)
\ee\\
In what follows we set $d=4$. Variation of the action (\ref{eq:redaction2}) with respect to $\Phi$ leads to the equation of motion (EOM)    
\be 
F'(\Phi)R^{(2)}+Z'(\Phi)(\partial\Phi)^2 +2 Z(\Phi)\nabla^b\partial_b\Phi-2V'(\Phi)=-2G\frac{\delta {\cal L}_M^{(2)}}{\delta \Phi}  
\label{eq:eom1}  
\ee
and variation with respect to $g^{\mu\nu}$ leads to the EOM 
\be 
\left[\nabla_\mu \partial_\nu-g_{\mu\nu}\nabla^a\partial_a \right]F(\Phi)+Z(\Phi)\left[\partial_\mu\Phi\partial_\nu\Phi-\frac{1}{2}g _{\mu\nu}(\partial\Phi)^2 \right]=g_{\mu\nu}V(\Phi)-2GT_{\mu\nu}^{(2)} 
\label{eq:eom2} 
\ee
\end{widetext}
Using the $2$-dimensional metric 
\be
ds^2=f(r)dt^2-f(r)^{-1}dr^2
\label{eq:dstd}
\ee
it is straightforward to show that the $2$-dimensional Ricci scalar is of the form
\be 
R^{(2)}=\frac{d^2f}{dr^2}
\label{eq:ricci}
\ee
Using  ${\cal L}_M^{(2)}=T=\rho^{(2)}-p_r^{(2)}$ \cite{Avelino:2018rsb}, eq. (\ref{eq:ricci}) and the ansatz $\Phi=r$ in eq.(\ref{eq:eom1}) we obtain the EOM
\be 
f''F' -2Zf'-Z'f-2V'=-2G(\rho'^{(2)}-p_r'^{(2)}) 
\label{eq:syst1}
\ee
where $\rho^{(2)}$ and $p_r^{(2)}$ are the 2-dimensional density and pressure respectively and the prime ($'$) denotes derivative with respect to $r$.

Also for $\mu=\nu=0$ in eq. (\ref{eq:eom2}) we obtain (with the same ansatz for $\Phi$)
\be
f'F'+2fF''+Zf-2V=-4G \rho^{(2)}  
\label{eq:syst2}
\ee
Similarly for $\mu=\nu=1$ eq. (\ref{eq:eom2}) gives 
\be
f'F'-Zf-2V=4Gp_r^{(2)}  
\label{eq:syst3}
\ee
The system eqs. (\ref{eq:syst1})-(\ref{eq:syst3}) is overdetermined since there is only one unknown function $f(r)$. Thus for a solution to exist eqs. (\ref{eq:syst1})-(\ref{eq:syst3}) must be equivalent to each other (up to a constant of integration). It may be shown that this consistency requires that 
\be
\begin{matrix}  
Z=-F''&&&&& \rho^{(2)}=-p_r^{(2)}
\label{eq:syst4}
\end{matrix}
\ee 
Indeed using eqs. (\ref{eq:syst4}), the system eqs. (\ref{eq:syst1})-(\ref{eq:syst3}) is equivalent to a single equation  
\be  
f'F'+fF''-2V=-4G \rho^{(2)}=4Gp_r^{(2)}  
\label{eq:sol1} 
\ee 

The general equation (\ref{eq:sol1}) connects the metric function $f$ with the geometric potential $V$ emerging from dimensional reduction and the  nonminimal coupling $F$ in the presence of a static spherically symmetric perfect fluid whose equation of state parameter is $-1$.  Thus,  any spherically symmetric metric of the form (\ref{eq:metric4d}) can emerge either due to an appropriate perfect fluid or as a vacuum solution of dimensionally reduced modified gravity with properly selected nonminimal coupling $F$ and/or potential $V$.

In what follows we focus on modifications of GR due to the geometric potential $V$ and fix $F$ to the GR form $F=\Phi^2$ implying $Z=-2$  (from eq. (\ref{eq:syst4})). Then  eq. (\ref{eq:sol1})  becomes 
\be 
rf'+f-V=-2G\rho^{(2)}=2Gp_r^{(2)}
\label{eq:solfl3} 
\ee
In order to quantify deviations from GR we set 
\ba
f(r)&=&1-g(r)
\label{eq:fg} \\ 
V(\Phi)&=&1+V_1(\Phi)
\label{eq:VV1} 
\ea
and expressing the dimensionally reduced density $\rho^{(2)}$ in terms of its $4$-dimensional counterpart $\rho$ as
\be 
\rho^{(2)}(r)=4\pi \Phi^2 \rho(r)
\ee 
in eq. (\ref{eq:solfl3}),  we obtain
\be 
\rho_{tot}(r)= \rho_m(r)+\rho_V(r)=\frac{1}{\kappa r}\left[\frac{g(r)}{r}+g'(r)\right]
\label{eq:reconstr}
\ee
where the geometric effective energy density is defined as
\be  
\rho_V(r)\equiv -\frac{V_1(\Phi)}{\kappa r^2}
\label{eq:rov}  
\ee
Therefore the generalization of the scalar-tensor potential leads to an effective energy density of geometric origin which generates the same spherically symmetric metric as a corresponding spherically symmetric perfect fluid with equation of state parameter $w=-1$ and energy density $\rho_m(r)=\rho_V(r)$. This derived equivalence between geometric and matter energy density allows the reconstruction of the geometric potential by demanding that its gravitational effects in the vacuum should be identical with the gravitational effects of a given matter fluid in the context of GR. This reconstruction from a realistic dark matter profile will be the main focus of the next section. 

\section{Special Cases - Reconstruction of Gravitational Action}
\label{sec:Reconstruction}
\subsection{Vacuum GR and Grumiller's gravity model} 

A special case of the geometric potential introduced in the previous section has been considered by Grumiller  \cite{Grumiller:2010bz,Grumiller:2011gg}. In particular, the following dimensionally reduced action was investigated
\be 
S=\frac{1}{4G}\int d^2x\sqrt{-g^{(2)}}[\Phi^2 R^{(2)}+2(\partial\Phi)^2+6\Lambda\Phi^2-8\alpha\Phi-2] 
\ee 
This is a special case of the general action (\ref{eq:redaction2}) with the GR coupling $F=\Phi^2$, $Z=-2$ and a geometric potential of the form
\be
V(\Phi)=1+4 \alpha \Phi -3 \Lambda \Phi^2
\label{eq:vgrum}
\ee
The ansatz $\Phi=r$ and our general reconstruction equation (\ref{eq:reconstr}) leads to the Schwarzschild-Rindler-deSitter metric function as a vacuum solution ($\rho_m=0$)
\be 
f(r)=1-2GM/r +2\alpha r - \Lambda r^2 
 \label{eq:fgrum}
\ee
in agreement with Grumiller's metric \cite{Grumiller:2010bz}.%\footnote{up to a - sign in quadratic geometric potential due to different metric signature.} The Schwarzschild term emerges as usual with the arbitrary integration constant $M$ while the constants $\alpha$ and $\Lambda$ (Rindler term and cosmological constant) are determined from the geometric potential.

The main advantages of the Grumiller potential (\ref{eq:vgrum}) include its simplicity and  its generic nature as it involves terms that dominate at large distances while at the same time it does not lead to any curvature singularities at infinity where the Ricci scalar (\ref{eq:ricci}) remains finite. On the other hand the metric function (\ref{eq:fgrum}) has been used to reconstruct the velocity profiles of galaxies without dark matter with mixed results \cite{Lin:2012zh,Mastache:2012ep,Cervantes-Cota:2013hza}. Even though it was found that the constant acceleration Rindler term can provide satisfactory fits to the observed velocity rotation curves of some galaxies in regions where these curves are rising with distance it became clear that for universal fits more parameters are needed in the potential. Such parameters however should be introduced in a way that is most efficient phenomenologically i.e. inspired from observed dark matter density profiles while at the same time they preserve the advantages of the Grumiller potential (simplicity and lack of large scale singularities). Using these principles in the next subsection we generalize the Grumiller geometric potential by demanding that the new potential reproduces in the vacuum the gravitational effects of a well known dark matter density profile parametrization: the Navarro-Frenk-White density profile \cite{Navarro:1995iw,Navarro:1996gj}. 

\subsection{Reconstruction of Geometric Potential}

The Navarro-Frenk-White (NFW) profile \cite{Navarro:1995iw,Navarro:1996gj} can give good fits to a wide range of observed rotation curves of galaxies in the context of general relativity (GR). It is of the form  
\be 
\rho_{NFW}(r)=\frac{\rho_o}{\frac{r}{R_s}(1+\frac{r}{R_s})^2}  
\label{eq:ronfw} 
\ee
where the scale radius $R_s$ and $\rho_o$ are parameters which vary from halo to halo.

The GR gravitational effects of this profile can be reproduced in the vacuum of a modified gravity model with a geometric  potential reconstructed using eq (\ref{eq:rov}) as 
\be 
\rho_V(r)\equiv -\frac{V_1(\Phi)}{\kappa r^2} =\rho_{NFW}(r)
\label{rhov-nfw}
\ee
which leads to a potential of the form
\be
V(\Phi)=1+\frac{4\alpha \Phi}{(1+\beta \Phi)^2}
\label{nfwpot}
\ee
with $\beta=\frac{1}{R_s}$ and $\alpha=2\pi G\rho_o R_s$. This potential reduces to the Rindler-Grumiller potential \cite{Grumiller:2010bz} for $\beta=0$. The new parameter $\beta$ introduces no large scale curvature singularities while it is designed to maximize the efficiency of fits to the observed rotation curves to the extend that such fit is obtained by the NFW density profile in the context of GR. Also the above potential reconstruction method can be easily generalized for any other density profile.

Solving eq. (\ref{eq:reconstr}) in the vacuum ($\rho_m=0$) with the geometric density $\rho_V$ obtained from the reconstructed potential (\ref{nfwpot}) we obtain the term $g(r)$ of the metric function 
\be 
g(r)= \frac{C}{r}- \frac{4\alpha[\frac{1}{1+\beta r}+ ln(1+\beta r)]}{\beta^2 r} 
\label{grwb}
\ee
where $C$ is a constant of integration.
Expansion of $g(r)$ of eq. (\ref{grwb}) as a power series demonstrates that this metric function is a generalization of the Rindler-Grumiller metric function (\ref{eq:fgrum}) for $\Lambda=0$
%\begin{widetext}
\be 
g(r)=\frac{C-\frac{4\alpha}{\beta^2}}{r}-2\alpha r+\frac{8}{3}\alpha\beta r^2  +O(r)^3
\ee
%\end{widetext} 
which after a redefinition of the integration constant $C$ clearly reduces to the Rindler-Grumiller metric function for $\beta r\ll 1$.
Setting  $C=2GM +\frac{4\alpha}{\beta^2}$ and using (\ref{eq:fg}),  the metric function $f(r)$ becomes 
\be 
f(r)=1-\frac{2GM}{r}-4\alpha \frac{1-\frac{1}{1+\beta r}- ln(1+\beta r)}{\beta^2 r}  
\label{eq:flp} 
\ee
which generalizes the Grumiller metric (\ref{eq:fgrum}) with one additional parameter ($\beta$) and is based on the geometric potential reconstructed from the NFW density profile. In the next subsection we check the efficiency of this metric in fitting two representative observed velocity rotation curves. The quality of fit will also be compared with the corresponding fit of of the Rindler-Grumiller metric \cite{Grumiller:2010bz}.

\subsection{Fitting Velocity Rotation Curves}

It is straightforward to show that the radial timelike geodesics in a backround metric of the form (\ref{eq:metric4d}) may be written as 
\be 
\frac{1}{2}\left(\frac{dr}{d\tau}\right)^2+V^{eff}=\frac{k^2}{2}  
\ee  
where $k$ is a constant,
\be 
V^{eff}=\frac{f(r)}{2}\left(1+ \frac{l^2}{r^2}\right) 
\ee
is the effective potential and  $l$ is the constant angular momentum per unit mass.

In the special case of the vacuum Schwarzschild-Rindler-deSitter metric function (\ref{eq:fgrum}) the effective potential reads
\be 
V^{eff}=-\frac{GM}{r}+\frac{l^2}{2r^2}-\frac{GMl^2}{r^3}-\frac{\Lambda r^2}{2}+\alpha r\left(1+\frac{l^2}{r^2}\right)
\label{eq:veffg}
\ee
In what follows we set $\Lambda=0$ since the effects of the cosmological constant can be ignored on galactic scales. For the  metric function (\ref{eq:flp}) emerging from the NFW reconstructed potential (\ref{nfwpot}) we have
\begin{widetext}
\be 
V^{eff}=-\frac{GM}{r}+\frac{l^2}{2r^2}-\frac{GMl^2}{r^3}-\frac{2\alpha}{\beta^2 r}\left[1-\frac{1}{1+\beta r}-ln(1+\beta r)\right]\left(1+\frac{l^2}{r^2}\right) 
\label{eq:vefflp}   
\ee  
\begin{figure}[!t]
\centering
\vspace{0cm}\rotatebox{0}{\vspace{0cm}\hspace{0cm}\resizebox{0.75\textwidth}{!}{\includegraphics{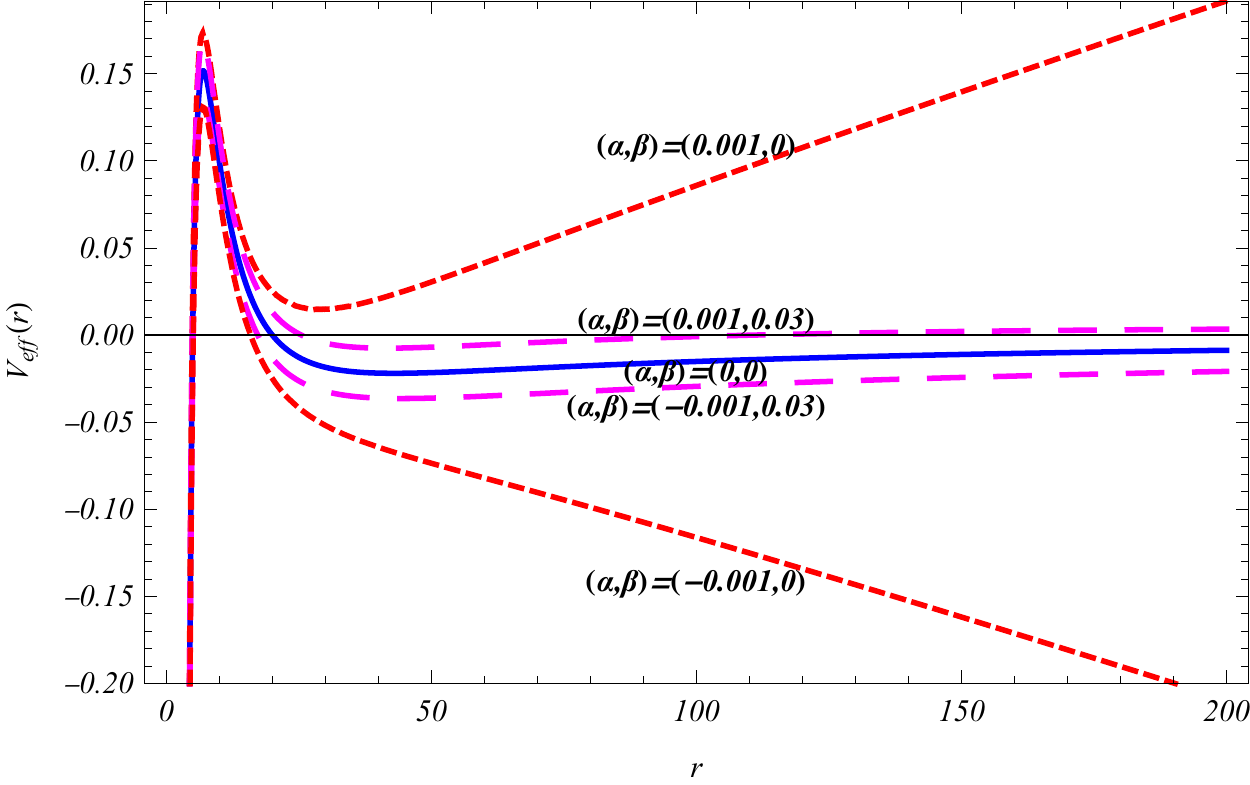}}}
\caption{The effective potential (\ref{eq:vefflp}) that determines the velocity rotation curves for parameter values $l=10$, $M=2$. The GR prediction (continous blue line) is obtained for $\alpha=0$ while the upper and lower red short-dashed lines correspond to the Rindler metric ($\beta=0$) with $\alpha>0$ and $\alpha<0$ respectively. The upper and lower pink long-dashed lines correspond to the metric of the reconstructed potential ($\beta>0$) for $\alpha>0$ and $\alpha<0$ respectively. In the later cases the GR prediction is obtained for large enough values of $r$.} \label{fig1}
\end{figure}
\end{widetext}
where we have dropped the constant terms on the RHS of eqs. (\ref{eq:veffg}) and (\ref{eq:vefflp}). A plot of this effective potential for various values of parameters is shown in Fig. \ref{fig1}.
The predicted rotation velocities of test particles may be approximated as 
\be 
\upsilon^2(r)\simeq r \vert\frac{\partial V^{eff}}{\partial r}\vert_{l=0} 
\label{v2general}
\ee
where we have set $l=0$ to avoid double counting of the angular momentum \cite{Cervantes-Cota:2013hza}.
Thus for the Schwarzschild-Rindler metric in the dark matter halo we have \cite{Grumiller:2010bz,Halilsoy:zva}\footnote{For further developments of this velocity profile see \cite{Mannheim:2005bfa,Mannheim:2010ti}} 
\be 
\upsilon^2(r)=\frac{GM}{r} +\alpha r
\label{rotvel1} 
\ee
where $M$ is the luminous mass in the galactic core. For the velocity profile corresponding to the NFW reconstructed potential (\ref{eq:vefflp}) we have
\begin{widetext} 
\be 
\upsilon^2(r)=\frac{GM}{r} +\frac{2\alpha}{\beta^2r}\left[1+\frac{\beta r-1}{1+\beta r}-\frac{\beta r}{(1+\beta r)^2}-ln(1+\beta r)\right]
\label{rotvel2}  
\ee
 \begin{figure*}[ht]
\centering
\begin{center}
$\begin{array}{@{\hspace{-0.10in}}c@{\hspace{0.0in}}c}
\multicolumn{1}{l}{\mbox{}} &
\multicolumn{1}{l}{\mbox{}} \\ [-0.2in]
\epsfxsize=3.3in
\epsffile{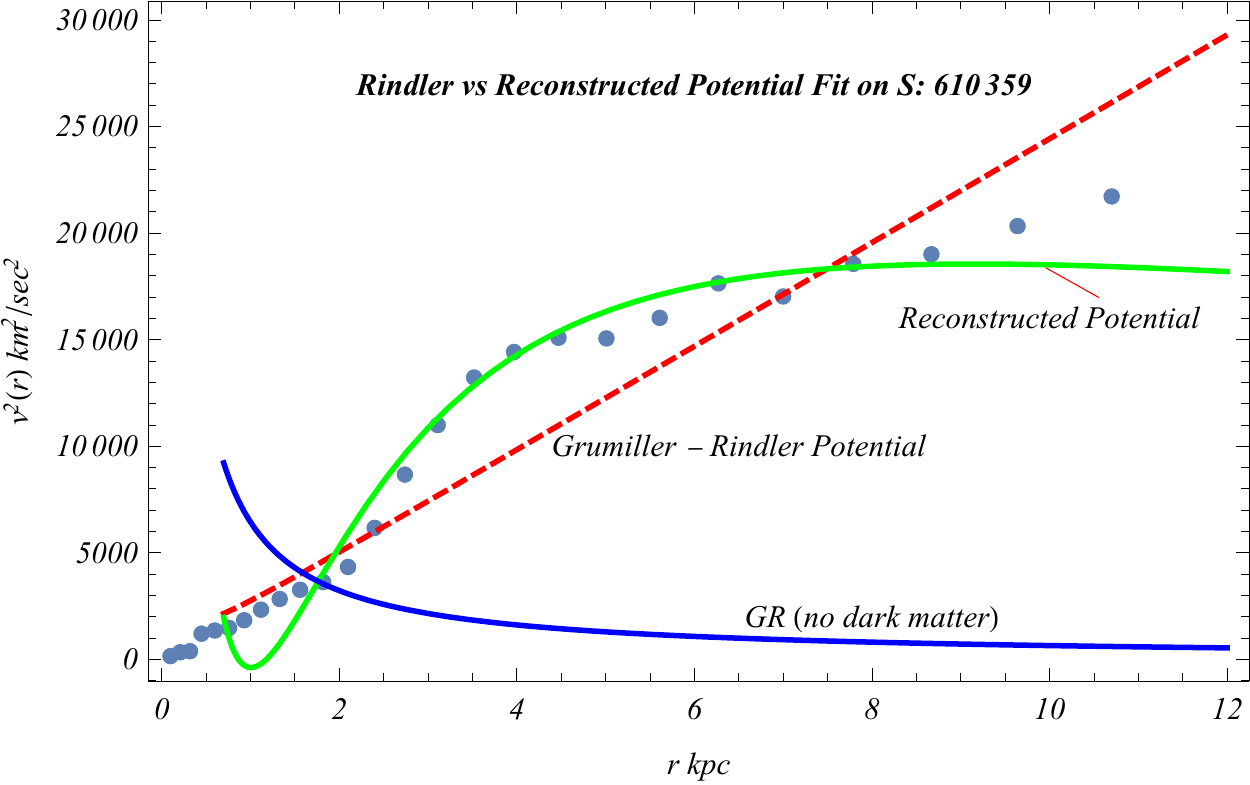} &
\epsfxsize=3.3in
\epsffile{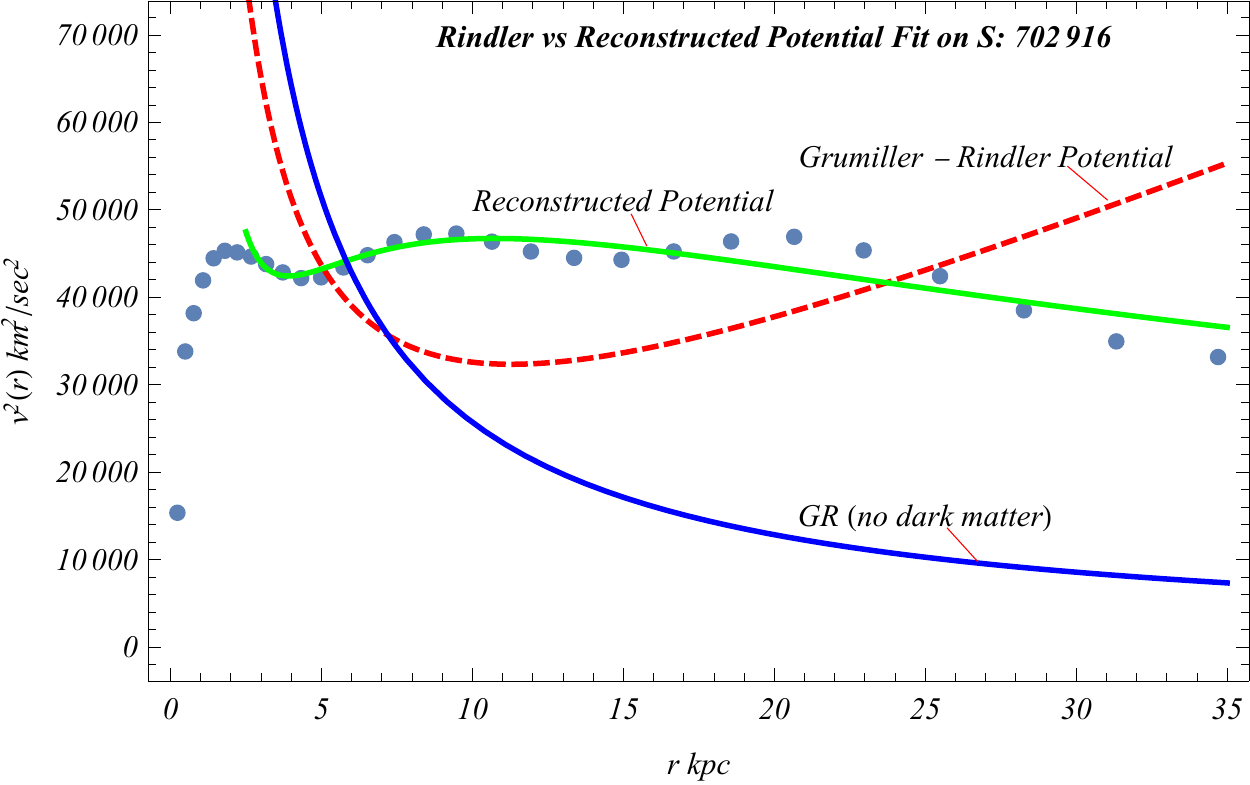} \\ 
\end{array}$
\end{center}
\vspace{0.0cm}
\caption{\small  The best fit forms of the velocity profiles (\ref{rotvel1}) (red dashed curve) and (\ref{rotvel2}) (green continuous curve) on the observed halo profiles (thick dots) of two typical galaxies (S:610359 left panel and S:702916 right panel). The blue continuous shows the fit of GR without dark matter which is clearly poor. }
\label{fig2} 
\end{figure*}
\end{widetext} 

The predicted rotation velocities (\ref{rotvel1}), (\ref{rotvel2}) can also be derived from the effective potentials (\ref{eq:veffg}) and (\ref{eq:vefflp}) assuming circular motion.  Setting
\be
\frac{dV^{eff}}{dr}=0
\label{veffeq0}
\ee 
solving (\ref{veffeq0}) for the angular momentum $l=\upsilon r$ and ignoring higher order terms. For example for the Grumiller effective potential (\ref{eq:veffg}) we obtain
\be 
\upsilon^2(r) \simeq \frac{GM}{r} +\alpha r+ 2GM\alpha +\frac{3G^2M^2}{r^2}-\alpha^2r^2
\label{v2sol}
\ee 
which reduces to (\ref{rotvel1}) if we ignore higher order terms in $M$ and $\alpha$.

The rotation curve is the sum of the following  three terms expressed by
\be
\upsilon^2(r)=\upsilon_G^2(r)+\upsilon_S^2(r)+\upsilon_{GM}^2(r)
\ee
where $\upsilon_G^2(r)$, $\upsilon_S^2(r)$ and $\upsilon_{GM}^2(r)$ are the different contributions in velocity of gas, stars and gravitational model (Rindler-Grumiller or recostructed potential) respectively. The term $\upsilon_{GM}^2(r)$ gives rise to the velosity rotation curves of galaxies without incorporating dark matter halo.  
We assume that the density (of gas and stars) drops to zero at $r_{min}$. Thus, in our analysis, we use data in the range $r_{min}<r<r_{max}$ and for a total mass $M$ we obtain the best fit forms of the velocity profiles corresponding the Rindler-Grumiller and reconstructed geometric potential.\\  

In Fig. \ref{fig2} we show the best fit forms of the velocity profiles (\ref{rotvel1}) (red dashed curve) and (\ref{rotvel2}) (green continuous curve) on the observed halo profiles (thick dots) of two typical galaxies (S:610359 left panel and S:702916 right panel). Velocity rotation data were obtained from the ’S-sample’ of Ref. \cite{Sofue}. 
The S:610359 \cite{Sofue:2015tsa} (also known as UGC 10359) has a typical rising velocity profile and is a SB(s)cdpec\footnote{A late-type barred peculiar spiral galaxy. It have well-developed, open, and knotty spiral arms with little or no bulge and without rings structures (see Ref.\cite{Buta:2015jsa} for morphology types of galaxies)} galaxy from Gassendi HAlpha survey of SPirals (GHASP) \cite{10.1111/j.1365-2966.2005.09274.x}.  The spiral galaxy S:702916 \cite{Sofue:2015tsa} (also known as UGS 2916) has a flat and slowly droping velocity profile and  is a Sab\footnote{An intermediate-type unbarred spiral galaxy  with tightly-wrapped spiral arms and a significant bulge (see Ref.\cite{Buta:2015jsa})} galaxy from early type galaxies survey \cite{Noordermeer:2007ux}.
 
Clearly the velocity profile corresponding to the reconstructed geometric potential provides a much better fit to the data for both observed velocity profiles and especially for the flat velocity profile. This is demonstrated quantitatively by the adjusted $R^2$ statistic \cite{doi:10.1002/sim.3429,doi:10.1080/00031305.2016.1256839,doi:10.1080/00220970109600656} which measures the quality of fit of a parametrization to a given set of data penalizing also for increased number of parameters. As shown in Table \ref{table:surveys}, the value of the adjusted $R^2$ is much closer to its optimal value $1$ in the case of the velocity profile corresponding the reconstructed geometric potential than the Grumiller Rindler potential or the simple Newtonian potential without dark matter.   In Table \ref{table:surveys} we also show the best fit values of parameters for each fitted velocity profile which in the case of Rindler potential agrees with previous studies \cite{Grumiller:2011gg,Carloni:2011ha,Cervantes-Cota:2013hza,Iorio:2010tp}. Notice that the best fir value of $\alpha$ for the reconstructed potential is $\alpha<0$ with is consistent with eq. (\ref{rhov-nfw}) and the fact that $\rho_{NFW}>0$.  
\begin{widetext}  
\begin{center} 
\begin{table}   
\centering 
\begin{tabular}{|c|c|c|c|c|c|c|c|}
\hhline{========}
&\multicolumn{3}{c|}{ }&\multicolumn{4}{c|}{ }\\
&\multicolumn{3}{c|}{Grumiller-Rindler Potential}&\multicolumn{4}{c|}{Reconstructed Potential}\\
&\multicolumn{3}{c|}{ }&\multicolumn{4}{c|}{ }\\
\hline
 &  & &  & & & &  \\
 Galaxy & $\alpha$       & $M$     &$R^2$   & $\alpha$   &$\beta$  & $M$   & $R^2$  \\
  &  & &  & & & &  \\
     & $[\times10^{-11}\frac{m}{s^2}]$  &$[\times10^{10}M_\odot]$  &  &$[\times10^{-9}\frac{m}{s^2}]$ &$[\times10^{-20}m^{-1}]$ &$[\times10^{10}M_\odot]$ &  \\
      &  & &  & & & &  \\
\hline  
\hline 
 &  & &  & & & &  \\
 $S: 610359$&$7.90\pm0.36$&$0.01\pm0.02$&$0.959$&$-4.10\pm0.16$&$3.17\pm0.13$&$0.32\pm0.02$&$0.983$\\
 &  & &  & & & &  \\
\hline 
&  & &  & & & &  \\
$S: 702916$&$4.64\pm0.55$&$4.11\pm0.47$&$0.923$&$-4.78\pm0.38$&$1.79\pm0.12$&$3.72\pm0.27$&$0.998$\\
 &  & &  & & & &  \\ 
 \hhline{========}  
\end{tabular} 
\caption{\small  The best fit values of parameters and the corresponding value of the adjusted $R^2$  of the velocity profiles (\ref{rotvel1}) and (\ref{rotvel2}) on the observed halo profiles  of two typical galaxies S:610359  and  S:702916 (rotation curve data obtained from \cite{Sofue}).}
\label{table:surveys}
\end{table} 
\end{center} 
\end{widetext}

\section{Discussion - Outlook}
\label{sec:Discussion}

We have used dimensional reduction of spherically symmetric gravity to construct a modified gravity model whose vacuum spherically symmetric metric has the same gravitational effects as the NFW dark matter density profile in GR. The model is a generalization of the Grumiller model whose vacuum spherically symmetric metric includes a Rindler term in addition to the standard Schwarzschild and cosmological constant terms. We have also shown that for any spherically symmetric perfect fluid with proper equation of state ($w=-1$) there is a modified gravity model, defined by a geometric potential, whose spherically symmetric vacuum metric is the same as the GR metric in the presence of the given fluid.

In particular we have shown that in order to reproduce the GR gravitational effects of the NFW density profile in the vacuum, the  reconstructed dimensionally  reduced geometric potential is of the form $V(\Phi)=1+4 \alpha \Phi/(1+\beta \Phi)^2-3 \Lambda \Phi^2$ where $\alpha$, $\beta$ are  parameters and $\Phi(r)$ is a field emerging from dimensional reduction. In the limit $\beta\rightarrow 0$ this geometric potential reduces to the Grumiller potential (\ref{eq:vgrum})  \cite{Grumiller:2010bz,Grumiller:2011gg}.

The reconstructed potential has the following interesting features:
\begin{itemize}
\item
It leads to a vacuum metric that provides significantly better fits to the velocity rotation profiles than the Grumiller linear potential term that leads to Rindler term in the vacuum metric.
\item
It leads to a vacuum metric that reduces to the GR vacuum on scales much larger than the $\beta^{-1}$ or the galactic scales. Thus on cosmological scales it is consistent with \lcdm. In contrast, the Grumiller-Rindler term is comparable to the cosmological constant on cosmological scales thus spoiling homogeneity and diverging from the standard \lcdm cosmic accelerating expansion.
\item 
Due to its non-polynomial form, it involves no IR curvature singularities while being distinct from the Grumiller potential thus demonstrating that this potential is not the only potential free from IR singularities. 
\end{itemize}

The cosmological effects of the model considered could be examined under the assumption of the existence of a large number of homogeneously distributed isotropic centers leading to large scale homogeneity in addition to isotropy around a single center (spherical symmetry). In such a physical setup, the geometric fluid density $\rho_V$ defined in eq. (\ref{rhov-nfw}) could be extended on cosmological scales as a homogeneous and isotropic fluid by replacing $r$ with the scale factor $a$ over the Hubble parameter $H_0$. Thus, on dimensional grounds the corresponding homogeneous geometric fluid would have an energy density scaling as
\be 
\rho_V(a)= -\frac{4 \alpha H_0}{\kappa a(1+\beta a/H_0)^2}
\label{geommatfluid}
\ee
where the Hubble parameter $H_0$ has been introduced on dimensional grounds. The derivation of eq. (\ref{geommatfluid}) has been heuristic and based mainly on dimensional analysis. A proper derivation would involve the detailed superposition of homogeneously distributed centers of isotropy and is beyond the goals of the present analysis. Nevertheless, the following comments can be made on this predicted geometric homogeneous dark matter
\begin{itemize}
\item
For $\beta=0$ the geometric fluid energy density reduces to the Rindler fluid whose energy density scales like $1/r$ or $1/a$ in a cosmological setup. This scaling is distinct from the matter density ($\sim 1/a^3$), the effects of spatial curvature ($\sim 1/a^2$) and the cosmological constant (constant effective density). Such a physically motivated and simple term can  be efficiently constrained using cosmological data probing the evolution of the Hubble parameter $H(a)$ even though a homogeneous component of ordinary dark matter would be required for a proper fit in addition to the cosmological constant.
\item
For $\beta \gg H_0$ which is expected for a value of $\beta$ reconstructed from galactic rotation curves, the geometric fluid density (\ref{geommatfluid}) scales as $1/a^3$ ie as ordinary homogeneous dark matter. Thus such a geometric fluid would not only provide better fits of galactic rotation curves but could also provide the homogeneous dark matter on cosmological scales. Such a geometric dark matter would have a predicted scaling signature of the form (\ref{geommatfluid}) leading to constraints on $\beta$ from both galactic rotation curve data and cosmological data probing the cosmic expansion rate. The consistency of these constraints could provide an efficient test for this class of models.
\end{itemize}

Other interesting extensions of our analysis include the following
\begin{itemize}
\item
The reconstruction of the geometric potential obtained from other special cases of spherically symmetric vacua. Such metrics could have oscillating components leading to oscillating terms in Newton's law at sub-mm scales which appear to be mildly favored by some short range gravity experiment data \cite{Perivolaropoulos:2016ucs,Antoniou:2017mhs}.
\item
The use of solar system data, short range gravity experiments data or other velocity profile data to impose constraints on the parameters $\alpha$ and $\beta$ of the reconstructed potential (\ref{nfwpot}).
\item
The generalization of the dimensionally reduced modified gravity model (\ref{eq:redaction2}) in different directions including a more general form of the nonminimal coupling (beyond $F(\Phi)=\Phi^2$), the consideration of $f(R^{(2)})$ extensions of the dimensionally reduced model or the generalization of the ansatz $\Phi = r$ used for the derivation of the spherically symmetric vacuum metric.
\end{itemize}
In conclusion, dimensional reduction in the context of spherical symmetry offers an interesting point of view for the modification of GR and can lead to a wide range of testable physically motivated models for gravity.
 
\section*{Acknowledgements}
We thank Daniel Grumiller for interesting and useful comments at the early stages of this project.\\

\textbf{Supplemental Material:} The Mathematica files used for the numerical analysis and for construction of the figures can be found in \cite{suppl}.\\

\bibliography{bibliography}              

\end{document}